\documentclass[
 reprint,
groupedaddress,
 amsmath,amssymb,
 aps,
twocolumn,
]{revtex4}

\usepackage{graphicx}
\usepackage{dcolumn}
\usepackage{bm}
\usepackage{xcolor} 
\usepackage[colorlinks=true,allcolors=blue,linkcolor=blue,citecolor=blue]{hyperref}
\usepackage{soul}

\newcommand{\Fref}[1] {Fig. \ref{#1}}

\newcommand{\Wcm}[2]{
$\rm {#1}\times10^{{#2}}~W/cm^2$}
\newcommand{\ba}{\begin{eqnarray}}
\newcommand{\ea}{\end{eqnarray}}
\newcommand{\br}{\begin{eqnarray*}}
\newcommand{\er}{\end{eqnarray*}}
\newcommand{\be}{\begin{equation}}
\newcommand{\ee}{\end{equation}}

\begin{document}

\title{High order harmonic generation in semiconductors \\ driven  at
  near- and mid-IR wavelengths}

\author{David Freeman}
\email{david.freeman@anu.edu.au}
\affiliation{Research School of Physics, Australian National
  University, Canberra, ACT, 2601, Australia}
\author{Shunsuke Yamada}
\affiliation{ Center for Computational Sciences, University of
  Tsukuba, Tsukuba 305-8577, Japan}
\author{Atsushi Yamada}
\affiliation{ Center for Computational Sciences, University of
  Tsukuba, Tsukuba 305-8577, Japan}
\author{Anatoli Kheifets}
\affiliation{Research School of Physics, Australian National
  University, Canberra, ACT, 2601, Australia}
\author{Kazuhiro Yabana}
\affiliation{ Center for Computational Sciences, University of
  Tsukuba, Tsukuba 305-8577, Japan}

\date{\today}

\begin{abstract}
We study high order harmonics generation (HHG) in crystalline silicon
and diamond subjected to near and mid-infrared laser pulses. We employ
time-dependent density functional theory and solve the time-dependent
Kohn-Sham equation in the single-cell geometry. We demonstrate that
clear and clean HHG spectra can be generated with careful selection of
the pulse duration. In addition, we simulate dephasing effects in a
large silicon  super-cell through a displacement of atomic positions
prepared by a molecular dynamics simulation. We compare our results
with the previous calculations by Floss ~et~al.  [Phys. Rev. A {\bf
    97}, 011401(R) (2018)] on Diamond at 800~nm and by Tancogne-Dejean
~et~al. [Phys. Rev. Lett. {\bf 118}, 087403 (2017)] on Si at 3000~nm.
\end{abstract}

\maketitle


\section{\label{sec:intro}Introduction}

High order harmonics generation (HHG) in solids has attracted efforts
of many experimental and theoretical groups. The ultimate goal of
these efforts is producing a compact and tunable source of coherent
XUV and soft X-rays on a chip that can be integrated into a
microelectronic device. The recent advances of HHG in solids has been
reviewed in the experiment \cite{Ghimire2019} and theory
\cite{Yu2019}. Previous efforts on HHG in condensed matter have been
driven by improved harmonic intensities offered by the higher atomic
densities of solids compared to those offered by atomic
gases. Experimental studies into HHG for thin films of bulk crystal
targets \cite{Ghimire2011, Schubert2014, Hohenleutner2015,
  Luu2015,Vampa2015,Langer2017,Liu2017,You2017,Shirai2018,Vampa2018,Orenstein2019}
have shown that improved intensities are achievable as demonstrated
theoretically \cite{Ghimire2019}. Targets have also recently been
expanded to structured metasurfaces \cite{Krasnok2018,Liu2018} and their 2D component
layers like graphene \cite{Yoshikawa2017,Yoshikawa2019,Hafez2018}.

Theoretical studies currently employ methods of varying quantitative
power to model the interaction of solid state targets with the pulsed
electric fields driving HHG. These methods include solution of
the time-dependent Schr\"odinger equation
\cite{Wu2015,Apostolova2018,Byun2021}, one-dimensional models
\cite{Hansen2017,Ikemachi2017,Bauer2018,Hansen2018,Jin2019}, density
matrix models \cite{Ghimire2012,Vampa2014,Vampa2015mark2} and
ab~initio descriptions like time-dependent density functional
theory (TDDFT)
\cite{Otobe2012,Floss2018,Floss2019,Otobe2016,Tancogne2017}. Recently, theories
have also been developed to describe HHG in more complex targets such
as bulk surfaces \cite{Georges2005,Golde2008,Breton2018} and
metasurfaces \cite{Shcherbakov2019}. Investigations using these
methods have revealed the contribution of inter- and intraband
mechanisms of the HHG process which are visualized in the energy band
picture \cite{Vampa2014,Ghimire2019}. Similarly, the impact of band
structure, band selection and laser polarization  have been
studied at length theoretically
\cite{Langer2017,You2017,Yoshikawa2019}.

Theoretical description of HHG in bulk crystalline targets requires
dealing with light propagation effects. Indeed, in the experiment, a
strong dephasing effect is often observed for condensed matter targets
\cite{Hohenleutner2015,Luu2015,Vampa2014,Luu2016,Yu2016,You2017}. This
effect has been explored in theoretical studies
\cite{Floss2018,Floss2019,Orlando2020} and found to be prominent in bulk solids
due to the higher atomic density than found in atomic gases. However, ultra-fast dephasing times of the order of
1~fs that has been proposed in theory \cite{Kim2019,Du2019} have posed
a problem for the strong-field community. Recently, these dephasing
times have been reconsidered through simulation of the linear response
of the current density to broadband excitation \cite{Floss2018}. This
analysis proposed dephasing times of the order of 10~fs and rectified
the experimental and theoretical spectra generated using TDDFT and
semiconductor Bloch equations (SBE) through simulation of the laser
focus propagation effect.

One of the major goals of theoretical investigations of HHG in solids
is to find the optimum conditions to generate clear and strong
harmonic signals.  In the Maxwell-SBE calculations \cite{Floss2018},
it was shown that the propagation effect is essential to obtain a
clear HHG signal of high order due to the dephasing effect.  Using the
single-scale Maxwell-TDDFT scheme \cite{Yamada2021}, an effort was
made to find the optimal thickness of a thin film that produces the
most intense HHG signals.  It was shown that a very thin film of
thickness of 2 - 15 nm is the optimum choice to produce intense HHG
signals.  In the present work, we investigate yet another unexplored
factor, a time duration of the pulse that produces clear and intense
HHG signals of high order. For longer driving pulses, more cycles
near the maximum field strength contribute to the HHG process with the
matter current getting closer to a quasi-periodic behavior thus contributing
to cleaner harmonic peaks.
%
%
For the purpose of our investigation, we study HHG in
crystalline diamond (Di) and Si utilizing the TDDFT implementation
within the computational SALMON framework \cite{Noda2019}.
We limit ourselves with the 
single cell (SC) calculations, which solve the time-dependent Kohn-Sham (TDKS)
equation for the electron dynamics in TDDFT. These calculations
consider the interaction between the electrons inside the bulk
crystal targets and a pre-specified pulsed electric field. 
Within this scheme, we also investigate a dephasing effect by
considering a large supercell with thermally distorted atomic positions.
Calculation of HHG signals from the supercell will clarify the
dephasing effect coming from electron-phonon coupling.

The paper is organized as follows: In Sec.  \ref{sec:Methods} we
describe the theoretical techniques for both time-dependent HHG and
dephasing calculations. In Sec. \ref{sec:Results}, we present
numerical results and examine the HHG spectra. We conclude in
Sec. \ref{sec:Conclusions} by outlining the future of this method
through extension to other condensed-matter systems.

\section{Methods}
\label{sec:Methods}

\subsection{Single unit-cell method}

The SC-TDDFT method considers the interaction of a bulk crystal unit
cell with a spatially uniform electric field
\cite{Bertsch2000,Otobe2008}. 
This method 
has been used in the study of
nonlinear and ultrafast dynamics 
\cite{Floss2018,Floss2019,Otobe2012,Otobe2016,Tancogne2017,Yamada2020,Uemoto2019,AYamada2019,Schultze2014,Wachter2014}. Utilizing
the dipole approximation, the electronic motion can be described
through the Bloch orbitals of the bulk,
$u_{n\bold{k}}(\bold{r},t)$, where the wave vector $\bm
k$ is contained within the three dimensional (3D) Brillouin zone of
the unit cell. The TDKS equation for such orbitals is written as:
%
\begin{align}
\label{eqn:2}
    i\hbar
    \frac{\partial{}u_{n\bold{k}}(\bold{r},t)}{\partial
      t} = \left\{\frac{1}{2m} \left( -i\hbar\nabla + \hbar \bold{k} +
    \hat{\bold{x}} \frac{e}{c} A(t)\right)^2 \right.\\ \left.
    - e\phi(\bold{r},t) + \delta \hat{V}_{ion} + V_{xc}(\bold{r},t) \right\} u_{n\bold{k}}(\bold{r},t).
\nonumber
\end{align}
%
The applied electric field in this approach is defined by the vector
potential of the incident pulse, $A(t)$. The electric potential
$\phi(\bold{r},t)$ includes the Hartree potential from the electrons
and the local contribution of the ionic potential. The terms
$\delta V_{\text{ion}}$ and $V_{xc}(r,t)$ describe the nonlocal part of the
ionic pseudopotential \cite{Troullier1991} and the
exchange-correlation potential \cite{Perdew1981}, respectively. In our
calculations, this exchange-correlation potential utilizes the
adiabatic local-density approximation (LDA) \cite{Onida2002}. 
Note that the LDA underestimates the bandgap of Si, as it does for diamond and other dielectrics. 
The experimental direct bandgap of Si is 3.4 eV while LDA  gives 2.4 eV.
We neglect the exchange-correlation term in the vector potential
$A_{xc}(t)$ for simplicity and do not rigorously model the
exchange-correlation effects of these types for infinite periodic
systems \cite{Ullrich2012,Vignale1996}.

The averaged electric current density, used in generating HHG spectra,
is calculated as follows:
%
\begin{align}
\label{eqn:3}
  \bm{J}(t) = &-\frac{e}{m} \int_{\Omega} \frac{d\bm{r}}{\Omega} \sum^{\text{occ}}_{n\bm{k}} u_{n\bm{k}}(\bm{r},t) \\
  &\times \left( -i\hbar\nabla + \hbar\bm{k} +
  \mathbf{\hat{x}}\frac{e}{c} A(t) \right)
  u_{n\bm{k}}(\bm{r},t) + \delta \bm{J}(t)
\ .
\nonumber
\end{align}
%
Here $\Omega$ is the volume of the unit cell and $n$ is the band index
restricted to occupied bands. The term $\delta J(t)$ denotes the 
contribution to the current density from the nonlocal part of the
pseudopotential \cite{Bertsch2000} and is given as follows:
%
\begin{align}
  \nonumber
  \delta \bm{J}(t) = &-\frac{e}{m} \int \frac{d\bm{r}}{\Omega}
  \sum^{\text{occ}}_{n\bm{k}} u^*_{n\bm{k}}(\bm{r},t)
  e^{-i\left[ \bm{k} + \bm{\hat{x}}(e/c) A(t) \right] \bm{r}}
  \\ &\times \frac{\left[ \bm{r}, \delta \hat{V}_{\text{ion}}
      \right]}{i\hbar} e^{i \left[ \bm{k} + \bm{\hat{x}} \frac{e}{c}
      A(t) \right] \bm{r} } u_{n\bm{k}}(\bm{r},t)
\label{eqn:4}
\end{align}

\subsection{Simulation details}
\label{sub:SimDetails}

We utilize the open-source software package SALMON (Scalable Ab initio
Light-Matter simulator for Optics and Nanoscience) \cite{Noda2019}. In
this code, a uniform 3D spatial grid is defined for the electron
orbitals. 
%
%
The time evolution of the electron orbitals is carried out using the
Taylor expansion method \cite{Yabana1996}.

For TDDFT calculations without dephasing, we define a cubic unit cell
of the diamond structure containing eight atoms with varying side
lengths depending on the specific target. For Di and Si calculations,
the unit cell dimensions are defined as $a_{\text{Di}} = 0.357$~nm and
$a_{\text{Si}} = 0.543$~nm. A spatial grid of $16^3$
points is used in Si and $24^3$ in Di calculations to enable an accurate description
of the Bloch orbitals. Convergence over spatial (r-point) grids is observed at these grid sizes. Similarly, the 3D Brillouin zone is sampled by
a grid made up of $32^3$ $k$-points. Note that convergence over the Brillouin zone is not observed at this grid size for diamond although qualitatively this size appears reasonable where we restrict ourselves to below the $50^{\text{th}}$ order harmonic. \citet{Otobe2012} finds similar grid parameters for diamond but with convergence observed over the Brillouin zone. The timestep is set for stability
of the calculation to be $2.5 \times 10^{-3}$~fs for Si and $1 \times 10^{-3}$~fs for Di. Note that tests
for convergence are important in these calculations due to the
discretization of the spatial, $k$-space and time parameters discussed
here.

For the Si calculation with dephasing, a supercell of
size $4^3$ describing a $512$ atom system is used. The Brillouin zone
is sampled by a grid made up of $8^3$ $k$-points.
Atomic positions are shifted from the equilibrium position with the diamond
structure while are frozen during the time evolution of electron orbitals.
They are generated from molecular dynamics simulations.
The detail of the preparation will be described below.

\subsection{Generating HHG spectra}

In our SC-TDDFT calculations, we utilize the following incident pulse
profile:

\begin{equation}
    \label{eqn:7}
    A(t) = -\frac{cE_0}{\omega} \text{sin}(\omega t)
    \text{cos}^6 \left[ \frac{\pi}{T} \left( t - \frac{T}{2} \right) \right], 
\ \ 
 0 \leqslant t \leqslant T
\ .
\end{equation}
Here $E_0$ is the peak amplitude of the electric field, $\omega$ is the
carrier frequency and $T$ is the full duration of the pulse. 
For the $\cos^6$ envelope, the FWHM duration $T_{\rm FWHM}$ is related to
the full duration T by $T_{\rm FWHM} \simeq 0.3 T$.
In our calculations, we consider the photon energies $\hbar \omega =
0.43$~eV ($\lambda=3000$~nm), $\hbar \omega =
0.62$~eV ($\lambda=2000$~nm) or $\hbar \omega = 1.55$~eV
($\lambda=800$~nm). We select typical pulse durations $T \gtrsim
100$~fs which are considerably larger than in previous HHG studies
($\sim32$~fs in \cite{Floss2018,Floss2019} and 30~fs in \cite{Yamada2021}).


In the SC-TDDFT formalism, the HHG spectrum is calculated as a Fourier
transform of the  matter current
\begin{align}
\mathrm{HHG}(\omega) &= \omega^2 | J(\omega)|^2, \label{HHGa} \\
J(\omega) &= \int dt e^{i\omega t} J(t) w(t), \label{HHGb} \\
w(t) &= 1 - \left(\frac{t}{T}\right)^2 + 2\times
\left(\frac{t}{T}\right)^3,
\label{HHGc}
\end{align}
where $w(t)$ is 
the window  smoothing function applied to remove spurious peaks in the spectra.
A Gaussian convolution of the form
\begin{equation}
    \mathrm{HHG_{av}}(\omega) = \int d\omega' \left( \pi \epsilon^2 \right)^{-1/2} \exp \left[ - \left( \frac{\omega - \omega'}{\epsilon^2} \right)^2 \right] \mathrm{HHG}(\omega'),
    \label{HHGd}
\end{equation}
is applied to carry out frequency averaging as often the case in experimental HHG measurements.


\section{Results}
\label{sec:Results}

\subsection{Bulk Si at mid-IR wavelengths}
\label{sub:mid-IRSi}

\begin{figure}[h]
    \includegraphics[width=90mm,height=70mm]{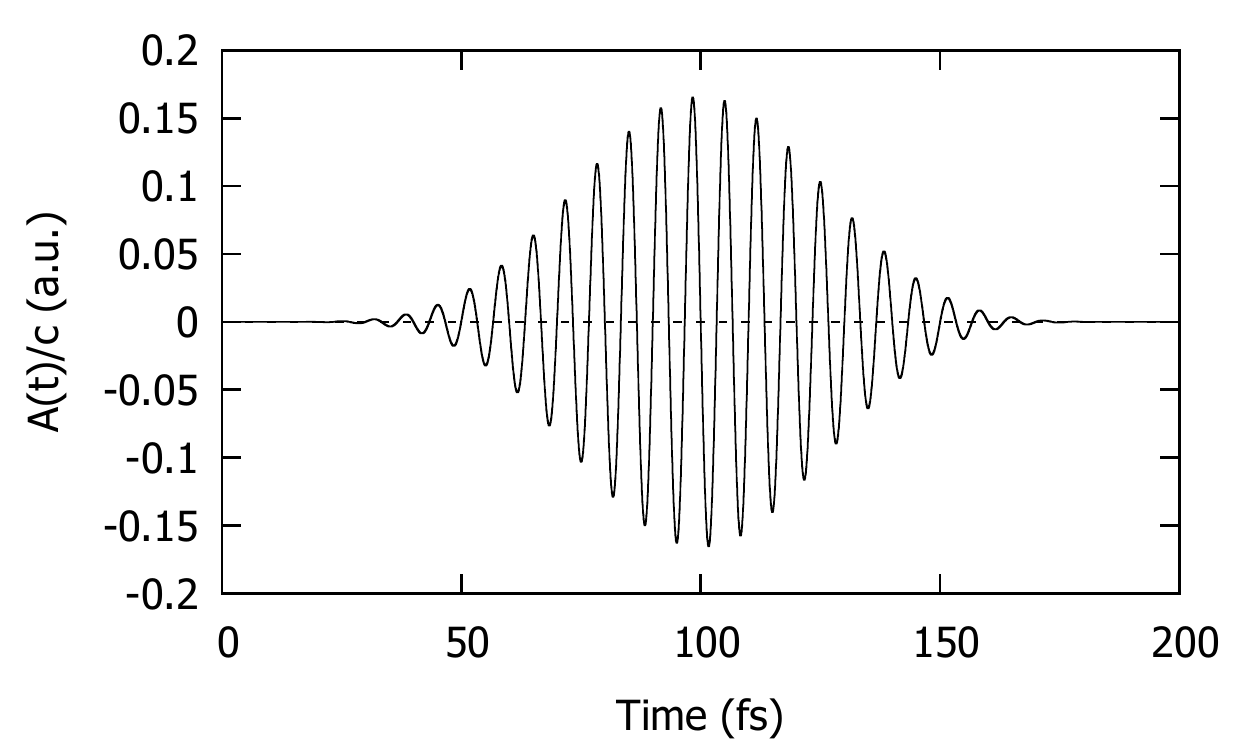}
    \bigskip
    \includegraphics[width=90mm,height=70mm]{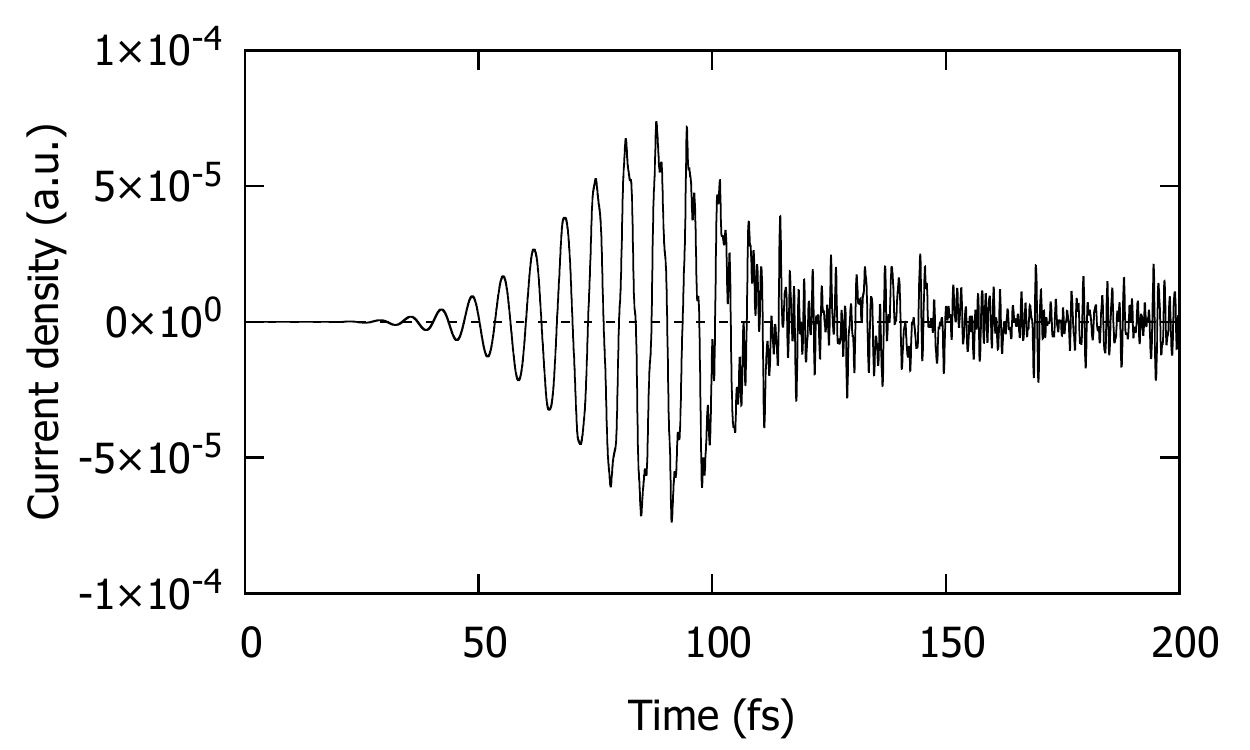}

    \caption{
    The applied vector potential (top)
    of a pulse with a full duration of $200$~fs at $\lambda=2000$~nm 
    and the peak intensity \Wcm{5}{11} and the
    matter current (bottom) in Si from a 
    SC-TDDFT simulation. 
     }
    \label{fig:currentordered}
\end{figure}

\begin{figure}[h]
    \includegraphics[width=90mm,height=70mm]{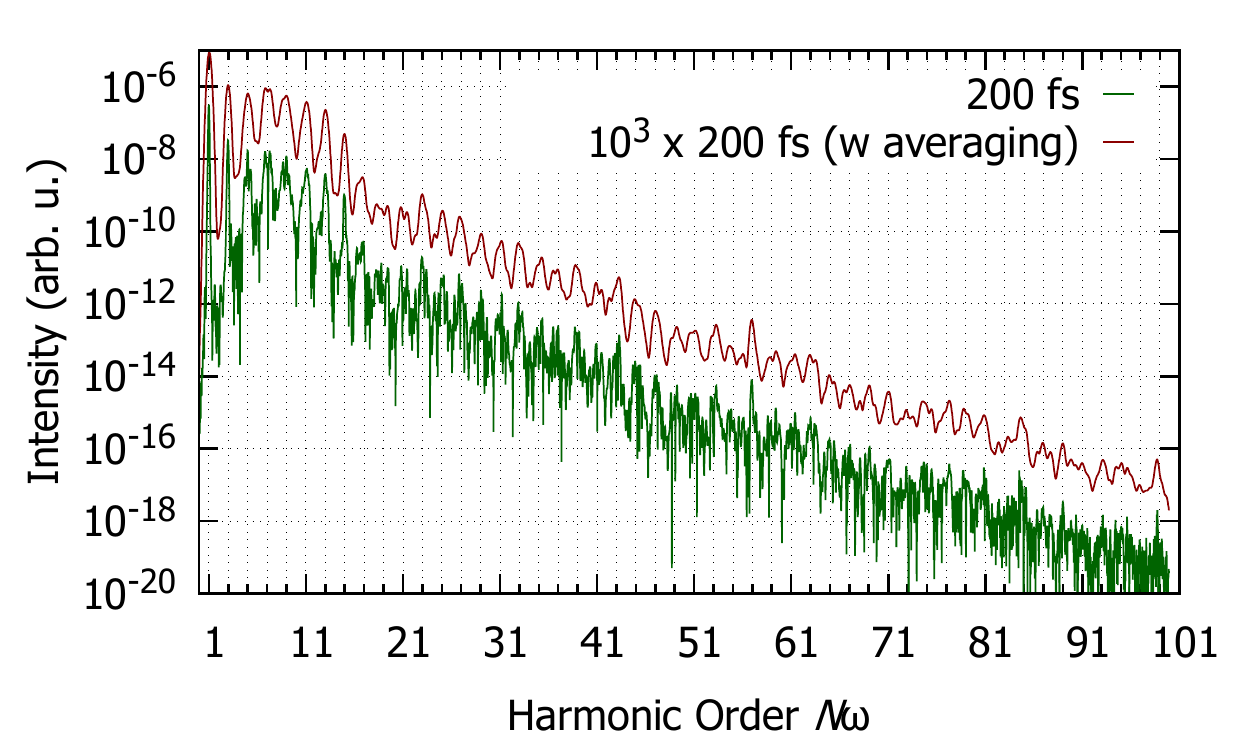}

    \caption{HHG spectra for Si are
      generated with a Fourier transform (green) and then Gaussian convolution (red) using Eq.~\ref{HHGd} with $\epsilon = 0.136$~eV. The spectra are the result of SC-TDDFT calculation considering interaction with a 
      pulse with a full duration of $200$~fs at $\lambda=2000$~nm and
      peak intensity \Wcm{5}{11}.
}
    \label{fig:fullFT}
\end{figure}

\begin{figure}[h]
    \includegraphics[width=90mm,height=70mm]{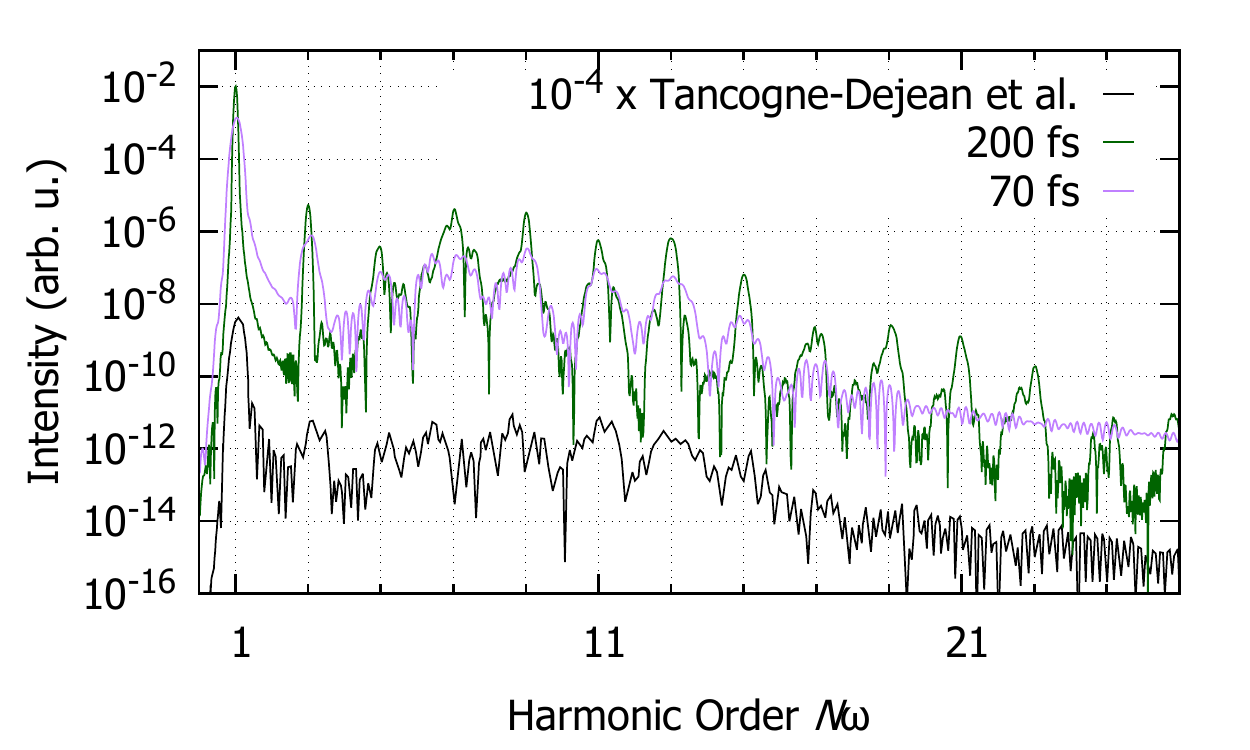}

    \caption{HHG spectra for Si are
      generated with a Fourier transform (green) and compared with calculation (purple) similar to \citet{Tancogne2017} (black). The spectra are the result of SC-TDDFT calculation considering interaction with a 
      pulse with full durations of $200$~fs and $70$~fs respectively, at $\lambda=3000$~nm and
      peak intensity \Wcm{1}{11}. Note that a $\cos^6$ envelope is applied here (for the pink calculation) instead of the $\cos^2$ carrier used by \citet{Tancogne2017}.
}
    \label{fig:TDcompare}
\end{figure}


We first consider bulk Si driven by a 2000 nm laser pulse with 200 fs full duration
with a peak intensity of \Wcm{5}{11}.
In ~\Fref{fig:currentordered}, the time profile  of the vector potential is shown
in the upper panel, and the induced matter current is shown in the lower panel.
Here the simulated matter current  displays large noise contributions that become particularly
prominent after around $110$ fs. Note also that the current is very asymmetric compared
to the symmetric driving pulse centered around $100$ fs.
We thus find that the SC-TDDFT method assuming a diamond structure
without any distortion effects records unphysically oscillating current
towards the end of the specified pulse length.
In Sec. III.C, we will show that the oscillating current mostly disappears once we introduce the dephasing effect caused by the thermal motion of atoms.

\Fref{fig:fullFT} displays the corresponding  
 HHG spectrum (in green) generated from the matter current shown in \Fref{fig:currentordered} 
 using Eqs.~(\ref{HHGa}-\ref{HHGc}).
We confirmed the simulation's convergence over spatial (r-point) and momentum (k-point) grids up to $51^{\text{st}}$ order harmonic for Si at 2000 nm. Nevertheless we show the spectrum up to $99^{\text{th}}$ order to demonstrate both the presence of, and our methods application to, higher order harmonics. Although HHG signals can be identified in the unaltered spectrum (green) even around the highest order
shown in the figure,
the harmonic structure appears noisy
and some harmonics cannot be clearly seen in this spectrum. 
Gaussian convolution using Eq.~(\ref{HHGd})
 overcomes this problem.
Figure \ref{fig:fullFT} shows the spectrum with the Gaussian filter (in red).

\Fref{fig:TDcompare} demonstrates the major effect that pulse duration has on the clarity and extent of HHG spectra, considering a $3000$~nm pulse interacting with Si. The $200$~fs calculation (in green) shows clear harmonic peaks beyond the $27^{\text{th}}$ harmonic order where the $70$~fs calculation generally shows unclear harmonics until the $15^{\text{th}}$ order. In addition to this, we note that the harmonic cutoff, comparing the simulations for the $200$~fs and $70$~fs pulses, appears dependent on pulse duration. We further examined this dependence through the application of an inverse Gabor transform focused to reveal the underlying current contributions associated with the highest order harmonics. Our analysis demonstrates, as expected, the importance of multi-cycle current accumulation in the formation of harmonic structure, where we observe that a $70$ fs, $3000$ nm pulse possesses too few cycles to capture the higher order returns.

Our Si results at 2000~nm (\Fref{fig:fullFT}) and 3000~nm (\Fref{fig:TDcompare}) need to be contrasted with the earlier
calculation on Si at 3000~nm presented by \citet{Tancogne2017}. In particular, using equivalent
parameters to those reported in \cite{Tancogne2017}, a direct comparison is achieved for 3000~nm in \Fref{fig:TDcompare}. \citet{Tancogne2017} recorded noisy harmonics of the lower order while their higher
order harmonics appeared clean. They related this effect with the
joint density of states (JDOS) as a measure of the overlap between the
valence and conduction bands. A large overlap and high JDOS would
promote the recombination and inter-band HHG. Somewhat
counter-intuitively, \citet{Tancogne2017} discovered
that a high JDOS was in fact detrimental to HHG emission making the
corresponding harmonic peaks noisy.  Conversely, those harmonics
falling into a low JDOS region of the HHG spectrum stayed clean. This
observation led the authors to conclude that it was the intra-band
mechanism that was largely responsible for HHG in silicon driven at
MIR. However, our HHG spectra at both $2000$~nm (\Fref{fig:fullFT}) and $3000$~nm (\Fref{fig:TDcompare}) at $200$~fs total duration are both entirely clean even without averaging. Note that we do not observe a clear and obvious JDOS effect in our calculations at both 2000~nm and 3000~nm when the pulse duration is sufficiently large.

\subsection{Bulk diamond  at near IR wavelengths}
\label{sec:Diamond}

\begin{figure}
    \includegraphics[width=90mm,height=70mm]{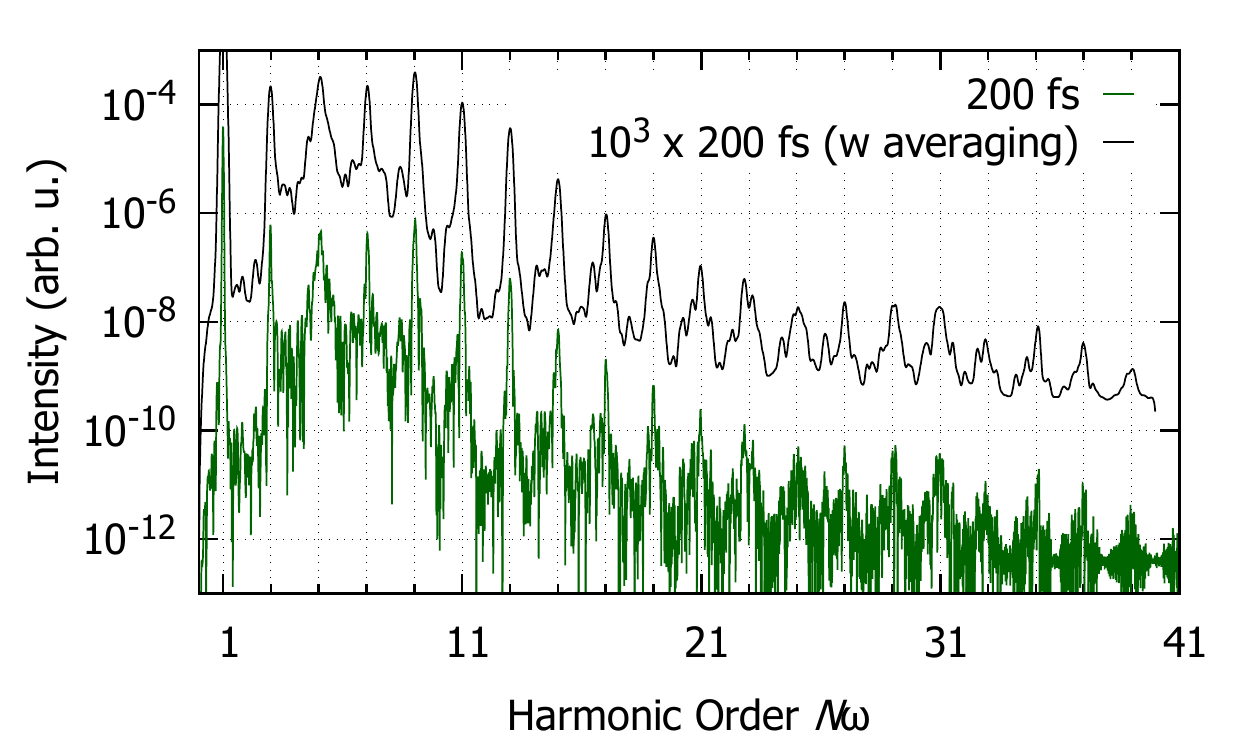}
\bigskip

    \includegraphics[width=90mm,height=70mm]{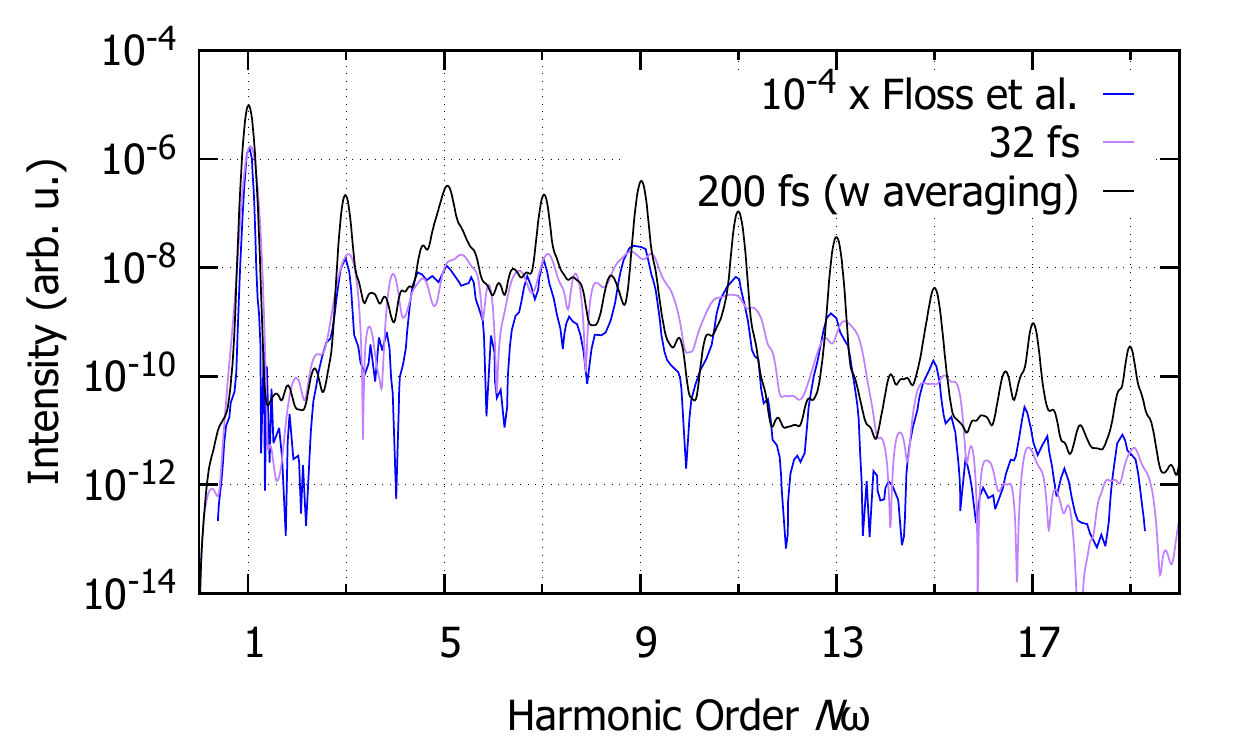}

    \caption{Top: The HHG spectrum of Di (shown in green) and Gaussian convoluted spectrum (in black), using Eq.~\ref{HHGd} with $\epsilon = 0.136$ eV, are generated with
      SC-TDDFT driven by a $800$~nm pulse with a full duration of
      $200$~fs and the peak intensity of \Wcm{1}{13}. Bottom: The analogous
      spectrum of \citet{Floss2019} with dephasing is displayed in blue and an SC-TDDFT
      calculation for a $32$ fs pulse is shown in purple. The top and
      bottom panels show the expanded and zoomed in low order HHG
      spectra, respectively.}
    \label{fig:800nm_diamond}
\end{figure}

In \Fref{fig:800nm_diamond} we display the HHG spectra of Di produced
by the SC-TDDFT. The primary calculation is driven by an $800$~nm incident
pulse of a $200$~fs full duration and the peak intensity of
\Wcm{1}{13}. The Gaussian convoluted spectrum again reveals clear harmonics up to
$41^{\text{st}}$ order of the
fundamental frequency $\omega$, shown in the top panel of
\Fref{fig:800nm_diamond}. The lower order harmonics of this spectrum
are zoomed in on the bottom panel.  The HHG spectrum is compared
with an analogous calculation of \citet{Floss2019}. 
The three sets of the SC-TDDFT calculations displayed in
\Fref{fig:800nm_diamond} are clearly differentiated by the pulse
duration.

These authors employed the SC-TDDFT driven by a 
\Wcm{2}{13} pulse with a total duration of $32$ fs \cite{Floss2018}.
Such a calculation did not produce a particularly clear HHG spectrum as is seen in
\Fref{fig:800nm_diamond}. It is only the Maxwell propagation technique
coupled with the SBE that allowed \citet{Floss2018} to generate a clear spectrum.  
We thus conclude that, with refined parameters and a longer pulse duration,
one can obtain distinguishable high harmonics within the SC-TDDFT alone, while in previous calculations \cite{Floss2019} the Maxwell propagation or an inclusion of the explicit dephasing effect was required to obtain clear spectra that are often observed in measurements.

\subsection{Dephasing Effect}

\begin{figure}
    \includegraphics[width=90mm,height=70mm]{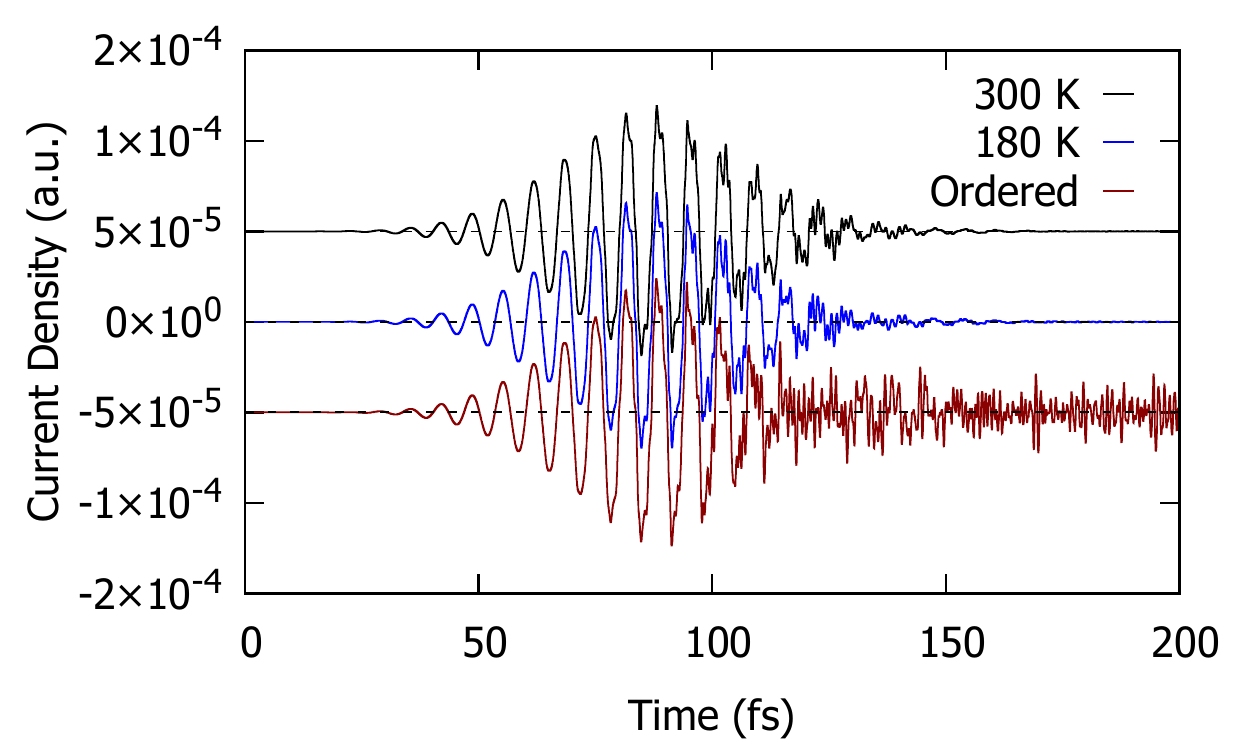}

    \caption{Examples of the dephasing effect  on the matter current
      in Si
      that arises from the application of a mean atomic position
      distortion. An effective temperature of $180$ K ($300$ K) is used in simulating the current density shown in blue (black). These examples consider a 200~fs full duration pulse
      impinging on bulk Si and are compared to current from a similar TDDFT calculation without distortion of atomic positions (red). A shift of $5 \times 10^{-5}$ is applied to help distinguish these results.}
    \label{fig:distortion}
\end{figure}

\begin{figure}[h]
\includegraphics[width=90mm,height=70mm]{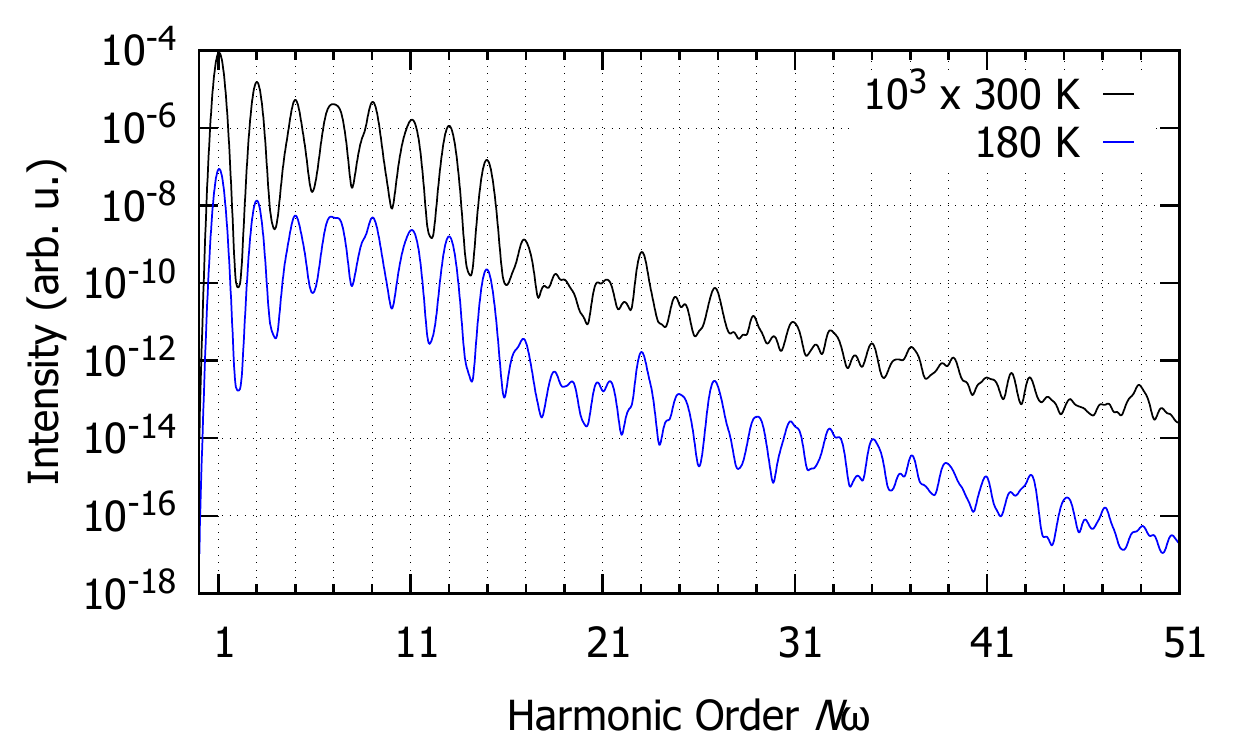}
\bigskip
\includegraphics[width=90mm,height=70mm]{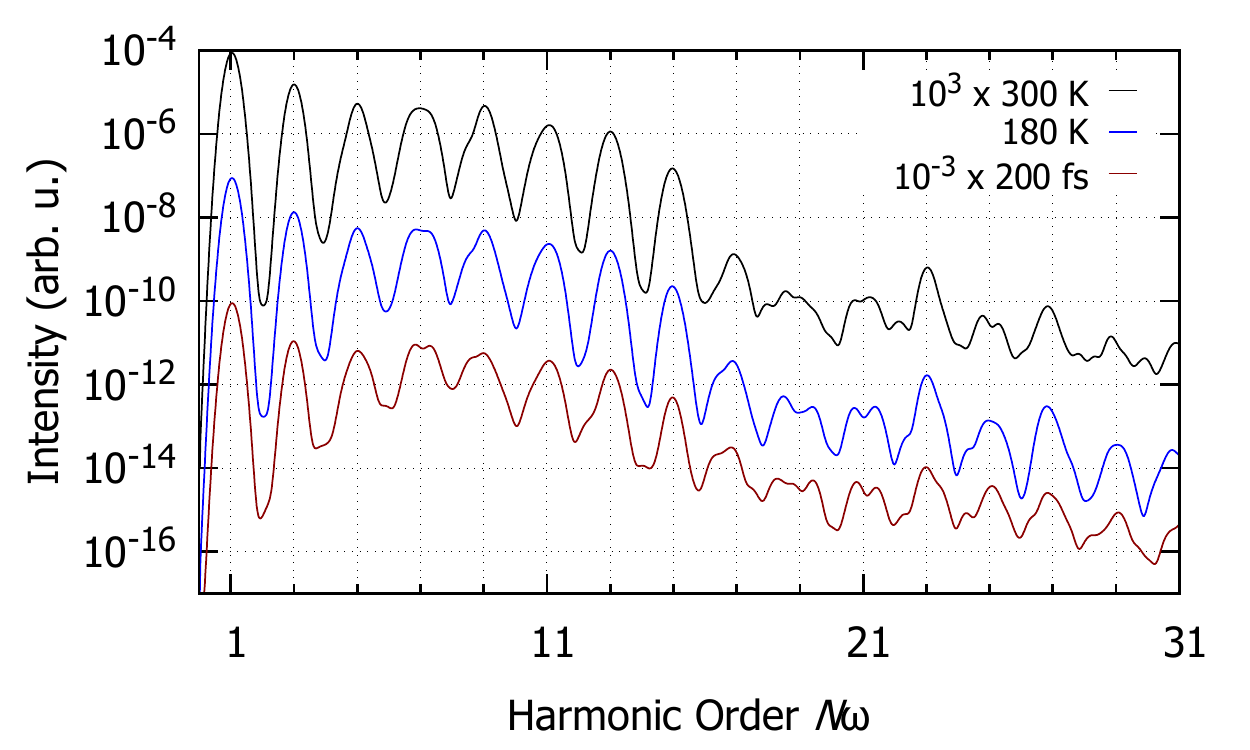}
    \caption{Comparison of convoluted HHG spectra from bulk Si with dephasing effect by an atomic displacement (top), together with its equivalent single-cell calculation (red, bottom) with Gaussian convolution applied using Eq.~\ref{HHGd} with $\epsilon = 0.136$ eV. Blue and black calculations here show the convoluted
    spectrum
    with atomic displacement by a thermal motion at $180$ K ($300$ K) in classical
    molecular dynamics calculation.}
    \label{fig:dephased_spectra}
\end{figure}

We expand the SC-TDDFT method to consider the atomic system at a
finite temperature by allowing the atomic positions to be displaced
from equilibrium. For this purpose, we consider bulk Si and first carry out a molecular
dynamics simulation with an empirical force field at a specified
temperature using a thermostat. We use LAMMPS software
\cite{Plimpton1995} to carry out the molecular dynamics simulation
where a three-body Stillinger-Weber potential \cite{PhysRevB.31.5262}
is used.  We pick up several atomic configurations with sufficiently
long intervals, and use these configurations in the SC-TDDFT. This represents a first-principles approach
introducing the dephasing effect into TDDFT for the first time without relying
on phenomenological parameter $T_2$. This is a convenient approach where we consider 
that previously ultrafast dephasing times $T_2 \sim 1$ fs had to be introduced 
to reconcile theoretical treatments and experiment.
During
the time evolution stage in the SC-TDDFT, the atomic positions
are fixed. We carry out calculations using supercells of different
sizes, and have found that results of different configurations are
quite similar to each other if we choose sufficiently large supercell.
We will  show below the calculation results  using a
supercell containing 512 atoms. Employing $8^3$ $k$-points, 
the total number of atoms is equal to $262,144$, that is
equal to the total number of atoms in a cell containing 8 atoms with
the $32^3$ $k$-points that we showed in
\Fref{fig:fullFT}. In fact, we numerically confirmed that calculations without
any distortion coincide with each other accurately between the two unit
cells, 8 atoms with $32^3$ $k$-points and 512 atoms with $8^3$
$k$-points.

We show electron current density in bulk Si under the applied electric field in \Fref{fig:distortion},
caused by the same applied pulse as shown in \Fref{fig:currentordered}.
The black (blue) calculation displays the current under the atomic configuration at a temperature of 300K (180K).
The matter current in disordered Si with
distortion  exhibits the desired dephasing after the pulse has ceased interacting.
With dephasing present, signals beyond $150$ fs appear significantly suppressed and the noise
contributions seen in \Fref{fig:currentordered} (red in \Fref{fig:distortion}) are removed.

Figure \ref{fig:dephased_spectra} demonstrates the effect of
incorporating dephasing into the HHG calculations in Si where Gaussian convolutions are also applied. 
The black calculation shows the HHG spectrum using
atomic positions from a molecular dynamics calculation at 300K.
While signals up to $100^{\text{th}}$ order can be seen in \Fref{fig:fullFT} without atomic distortion, here
we can see HHG signals up to $31^{\text{st}}$ order, and higher order signals
cannot be identified. We also note that the HHG signals themselves
become deeper between peaks and cleaner than those without atomic distortion, even when not incorporating a Gaussian convolution.
Therefore, the dephasing effect modified the HHG spectrum in two
ways, removing signals of HHG signals higher than $31^{\text{st}}$ order, and
increasing peak-to-peak depth for those signals that survive.

In Fig. \ref{fig:dephased_spectra}, the blue calculation shows the HHG spectrum using
atomic positions from a molecular dynamics simulation at 180~K.
Now we find HHG signals up to the $49^{\text{th}}$ order, where higher
order signals are suppressed.
This result at $180$ K clearly indicates that it is important to keep the material
cold in measuring HHG to observe higher order signals.
We however note that the distortion of atomic positions is generated
by the classical molecular dynamics calculation. At very low temperatures,
quantum zero-point motion will contribute significantly and should be
included in discussing the HHG spectra.

The results with the effective thermally-induced dephasing
effect, incorporated through atomic positions
generated by molecular dynamics simulations,
suggest that sample cooling could be used to greatly improve
the high-order harmonic signal intensity observed in experiment. In
principle, this should not pose a problem to current experiment as
liquid nitrogen or helium cooling is commonly utilized in a variety of condensed matter
physics contexts. 

It should be noted that we have opted to consider fixed ionic coordinates in our dephasing simulations instead of performing evolution through classical mechanics. We do not incorporate an Ehrenfest molecular dynamics approach into TDDFT as we consider that spurious coherence may occur in the classical approximation when considering solids as, when one considers molecules, the coherence between two states will soon disappear as the ionic wave packets evolve differently between ionized and non-ionized electronic states and the spurious coherence that may occur in molecular simulations could extend to simulations concerning solids.

\section{Concluding Remarks}
\label{sec:Conclusions}

We have presented here clear HHG spectra for diamond-like semiconductors 
interacting with near- and mid-IR pulsed electric fields.
We obtain clear harmonic spectra for sufficiently long pulses with durations $T_{\text{FWHM}} \gtrsim 70$~fs and these spectra are further clarified through Gaussian convolution.
We therefore review the
suggestions made in previous literature considering diamond
\cite{Floss2018,Floss2019} and silicon \cite{Tancogne2017}.  In crystalline Si,
we do not observe the expected joint density of states effect but do demonstrate a 
prominent dephasing effect on the HHG spectra. Using a first-principles approach introducing dephasing into TDDFT we
observe that higher-order harmonics are strongly suppressed by the
decoherence effects in the target whilst lower order harmonics appear
clearer in the spectra. On the basis of the strength of the
decoherence effects for Si at $2000$~nm laser wavelengths, we suggest
that helium or liquid nitrogen cooling of the target will improve higher order harmonic
intensity returns.

\section{Acknowledgements}
\label{sec:Acknowledgements}
The authors would like to acknowledge the two super computer resources
employed for TDDFT calculations: the National Computational
Infrastructure (NCI Australia) and 
Fugaku at RIKEN R-CCS, Japan, with
support through the HPCI System Research Project (Project ID:
hp210137).
This research was partially supported by JST-CREST under Grant No.
JP-MJCR16N5, by MEXT Quantum Leap Flagship Program (MEXT Q-LEAP)
under Grant No. JPMXS0118068681, and by JSPS KAKENHI Grant No.
20H02649.


\bibliography{apssamp_included.bib}


\providecommand{\noopsort}[1]{}\providecommand{\singleletter}[1]{#1}%

\end{document}